\documentclass[sensors,article,submit,pdftex,moreauthors]{Definitions/mdpi} 
\firstpage{1} 
\pubvolume{1}
\issuenum{1}
\articlenumber{0}
\pubyear{2026}
\copyrightyear{2026}
\datereceived{ } 
\daterevised{ } 
\dateaccepted{ } 
\datepublished{ } 

\Title{Performance Characterization of a Plastic-Scintillator Sensor for Fast-Neutron, Thermal-Neutron, and Gamma-Ray Discrimination}

\Author{Yuhang Liu $^{1}$, Fengpeng An $^{1,*}$, Guang Luo $^{3,*}$, Wei Wang $^{1,2,*}$, Xuesong Zhang $^{1}$, Dixiao Lu $^{1}$ and Xiaohao Yin $^{1}$}

\AuthorNames{Yuhang Liu, Fengpeng An, Guang Luo, Wei Wang, Xuesong Zhang, Dixiao Lu and Xiaohao Yin}

\address{%
	$^{1}$ \quad School of Physics, Sun Yat-sen University, No. 135 West Xingang Road, Guangzhou 510275, China; anfp@mail.sysu.edu.cn (F.A.); wangw223@mail.sysu.edu.cn (W.W.)\\
	$^{2}$ \quad Institut Franco-Chinois de l'Énergie Nucléaire, Sun Yat-sen University, No. 2 Daxue Road, Zhuhai 519082, China\\
	$^{3}$ \quad School of Science, Sun Yat-sen University, No. 66 Gongchang Road, Shenzhen 518107, China; luog7@mail2.sysu.edu.cn (G.L.)}

\corres{Correspondence: anfp@mail.sysu.edu.cn (F.A.); luog7@mail2.sysu.edu.cn (G.L.); wangw223@mail.sysu.edu.cn (W.W.)}

\abstract{Discrimination of fast neutrons, thermal neutrons, and $\gamma$ rays in mixed radiation fields is important for radiation monitoring, reactor-related measurements, and background suppression in nuclear experiments. In this work, we investigate a compact plastic-scintillator sensor composed of EJ276 or EJ200 optically coupled to an EJ426 thermal-neutron screen and read out by a single photomultiplier tube (PMT). The $\gamma$-equivalent energy response of the detector assemblies was calibrated using $^{137}$Cs, $^{22}$Na, and $^{60}$Co sources through Compton-edge analysis, and pulse shape discrimination was evaluated with an Am--Be neutron source under different moderator thicknesses. The EJ200+EJ426 assembly provides a well-separated discrimination between thermal-neutron capture events and $\gamma$-dominated events over the measured range, with a figure of merit greater than 5. In contrast, the EJ276+EJ426 assembly produced three identifiable signal populations associated with $\gamma$ rays, fast neutrons, and thermal neutrons. These results show that the proposed sensor architecture is a promising compact approach for mixed-field radiation applications.
}


\begin{document}


\section{Introduction}
Discrimination of $\gamma$ rays, fast neutrons, and thermal neutrons in mixed radiation fields is important for radiation monitoring, reactor-related measurements, shielding assessment, and background suppression in nuclear experiments~\cite{Bernstein:2019hix,Haghighat:2018mve,Gamage2015,Li:2022wqc}. Hybrid scintillation detectors that combine complementary response functions offer a practical route toward compact sensors for mixed radiation fields~\cite{ref-phoswich-review,ref-phoswich-ej276-gs20,Wilhelm:2017nima}. In particular, coupling a plastic scintillator to a thermal-neutron-sensitive screen can provide a simple single-readout architecture for a compact gamma--neutron detector~\cite{ref-phoswich-ej276-gs20,Wilhelm:2017nima,Griffiths:2020mlst,EljenEJ426}.

Recent progress in neutron--gamma discrimination has been driven by the development of pulse-shape-discriminating plastic scintillators, among which EJ-276 has attracted considerable attention because it combines the practical advantages of solid scintillators with fast-neutron/$\gamma$ separation capability~\cite{ref-ej276-calibration}. Previous studies have reported its $\gamma$ calibration, response functions, proton light output, and detection efficiency, indicating that EJ-276 is a suitable candidate for compact mixed-field detectors~\cite{ref-ej276-response}. For thermal-neutron detection, $^{6}$LiF/ZnS(Ag)-based screens such as EJ-426 provide an efficient conversion mechanism with relatively low $\gamma$ sensitivity, making them attractive as thermal-neutron-sensitive layers in composite detectors~\cite{EljenEJ426,GAMAGE20141,Perrey2021}. Notably, previous studies have shown that Phoswich detectors based on pulse-shape-discriminating scintillators and lithium-containing neutron-sensitive components can simultaneously resolve $\gamma$ rays, thermal neutrons, and fast neutrons with improved discrimination performance compared with single scintillators~\cite{ref-phoswich-ej276-gs20}.

In this work, EJ200, EJ276, and the $^{6}$LiF/ZnS(Ag)-based EJ426 screen were used to construct two compact detector assemblies, EJ200+EJ426 and EJ276+EJ426, for mixed-field response characterization. The detector concept is related to a phoswich-like configuration~\cite{ref-phoswich-review}, in which scintillators with different decay characteristics are optically coupled to a single photodetector and distinguished through pulse-shape analysis.

\section{Materials and Methods}
\subsection{Detector Configuration and Materials}
\label{sec:2.1}
\begin{figure}[H]
\includegraphics[width=10.0 cm]{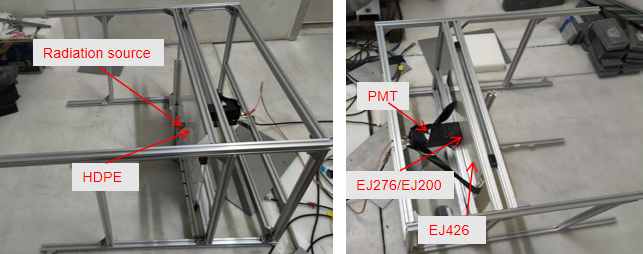}
\caption{Side view (left) and top view (right) of the experimental setup. The PMT is wrapped with black
tape and coupled with a plastic scintillator.}
\label{figswt}
\end{figure}

\begin{figure}[H]
\includegraphics[width=12.0 cm]{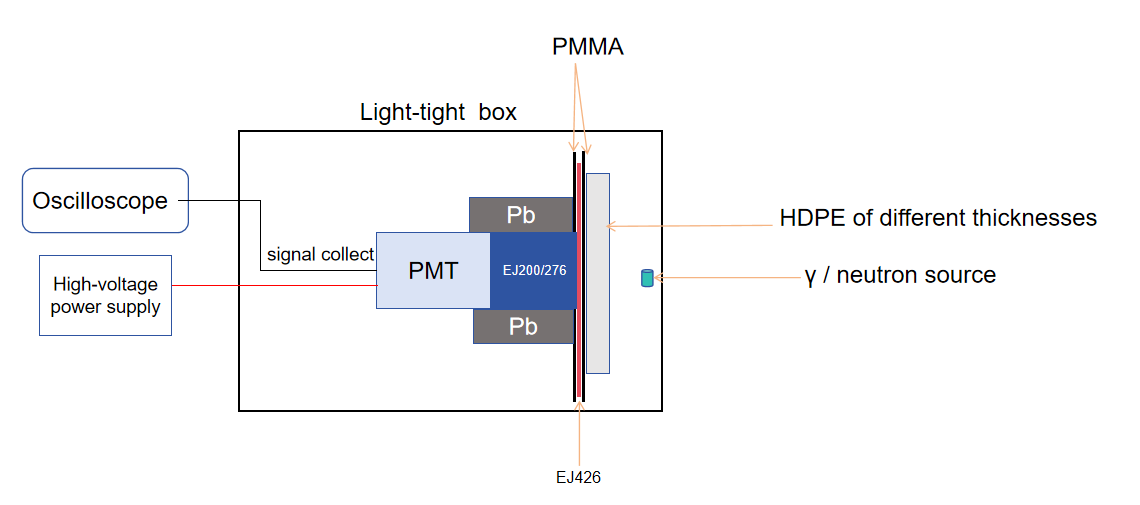}
\caption{Schematic diagram of the detector configuration. The detector assembly is enclosed in a light-tight box. The plastic scintillator is coupled to the PMT on one side and to the EJ426 neutron screen on the other side. The EJ426 screen is sandwiched between two PMMA plates, one of which has a $6.1~cm \times 6.1~cm$ opening. HDPE layers of different thicknesses are used for neutron moderation in the Am--Be measurements.
\label{fig1}}
\end{figure}

Figure~\ref{figswt} presents a photograph of the experimental setup and Figure~\ref{fig1} shows a schematic diagram of the detector configuration and operating principle. Signals were acquired using a digital oscilloscope (LeCroy HDO4104A) with a 1 GHz bandwidth and a sampling rate of 10 GS/s~\cite{HDO4104A}. The detector assembly was supported by an external metal frame and enclosed in a light-tight box made of thick black cloth to suppress ambient light interference. Optical coupling between the scintillator and the PMT was achieved using SL612 optical silicone grease~\cite{Haotangxing:SL612} (Beijing Hoton Technology Co., Ltd.). The scintillator modules were wrapped with aluminum foil to improve light collection. Lead bricks were placed around the plastic scintillator to reduce the environmental $\gamma$-ray background.


Because the EJ426 neutron screen is thin and fragile, it was mechanically protected by sandwiching it between two transparent polymethyl methacrylate (PMMA) plates. A square opening of $6.1~cm \times 6.1~cm$ was machined in the center of one PMMA plate to allow direct coupling between the plastic scintillator and the EJ426 neutron screen. On the source-facing side, the PMMA plate was attached to high-density polyethylene (HDPE) layers of different thicknesses for neutron moderation; this HDPE layer was not used in the $\gamma$-source experiments. Depending on the measurement purpose, either a neutron source or a $\gamma$-ray source was placed on the opposite side of the detector assembly.

The three scintillation materials used in this study were EJ200, EJ276, and EJ426. The basic properties of these materials are summarized in Table~\ref{tab:scintillator_properties}.
\begin{table}[H]
\caption{Basic properties of the scintillation materials used in this study.\textsuperscript{a}\label{tab:scintillator_properties}}
\begin{tabularx}{\textwidth}{CCC}
\toprule
\textbf{Material} & \textbf{Geometry/Dimensions} & \textbf{Emission Peak / Decay Time} \\
\midrule
EJ200 & Cube, 6~cm side length & 425~nm / 2.1~ns \\
EJ276 & Cube, 6~cm side length & 450~nm / $\gamma$: 13, 35, 270~ns; neutron: 13, 59, 460~ns\textsuperscript{b} \\
EJ426 & Screen, 42~cm$\times$42~cm $\times$ 0.32~mm & 450~nm / 200~ns \\
\bottomrule
\end{tabularx}
\noindent{\footnotesize{\textsuperscript{a}Data compiled from Eljen Technology datasheets and related references~\cite{EljenEJ200,EljenEJ276,EljenEJ426}.}}

\noindent{\footnotesize{\textsuperscript{b} Approximate mean decay times of the first three components for $\gamma$- and neutron-induced pulses~\cite{EljenEJ276}.}}
\end{table}

EJ200 and EJ276 were both plastic scintillator cubes with a side length of 6~cm and high light output. The EJ426 screen was a $^{6}$LiF/ZnS(Ag)-based thermal-neutron-sensitive screen with a thickness of 0.32~mm. EJ276 can distinguish between $\gamma$ rays and fast neutrons on the basis of their different pulse decay characteristics, whereas EJ200 has a faster decay time but does not provide pulse shape discrimination between these two components. EJ426 is suitable for thermal-neutron detection because it produces relatively slow scintillation signals and exhibits low $\gamma$ sensitivity in the present thin-screen configuration. Thermal neutrons are captured in EJ426 primarily through the $^{6}$Li$(n,\alpha)^{3}$H reaction shown in Equation~\ref{eq:li_capture}.
\begin{equation}
^{6}\mathrm{Li} + n \rightarrow \, ^{3}\mathrm{H} + \, ^{4}\mathrm{He} + 4.78~\mathrm{MeV}.
\label{eq:li_capture}
\end{equation}
This reaction produces $^{4}$He and $^{3}$H ions that subsequently excite the ZnS(Ag) phosphor and generate visible photons~\cite{Pino:2015dna,Stedman:1960}. The resulting light signals from EJ426 exhibit waveform characteristics that differ from the $\gamma$-induced signals in the plastic scintillator, thereby enabling thermal-neutron tagging in the composite detector. Therefore, EJ426 serves as an effective thermal-neutron-sensitive layer in mixed radiation fields.

For this experiment, an XP3232 PMT manufactured by Hainan Zhanchuang Photonics Technology Co., Ltd. ~\cite{Jiang:2022vnd} was employed. The XP3232 is a cylindrical vacuum tube with a length of approximately 11.2~cm and a diameter of approximately 5.1~cm. The PMT has a peak spectral sensitivity near 420~nm, which matches the emission spectrum of the plastic scintillator. The photocathode type was bialkali, and the effective photocathode diameter was approximately 5~cm. The PMT was coupled directly to the scintillator.

\subsection{PMT Gain Calibration}
\label{sec:2.2}
\begin{figure}[H]
	\includegraphics[width=12.0 cm]{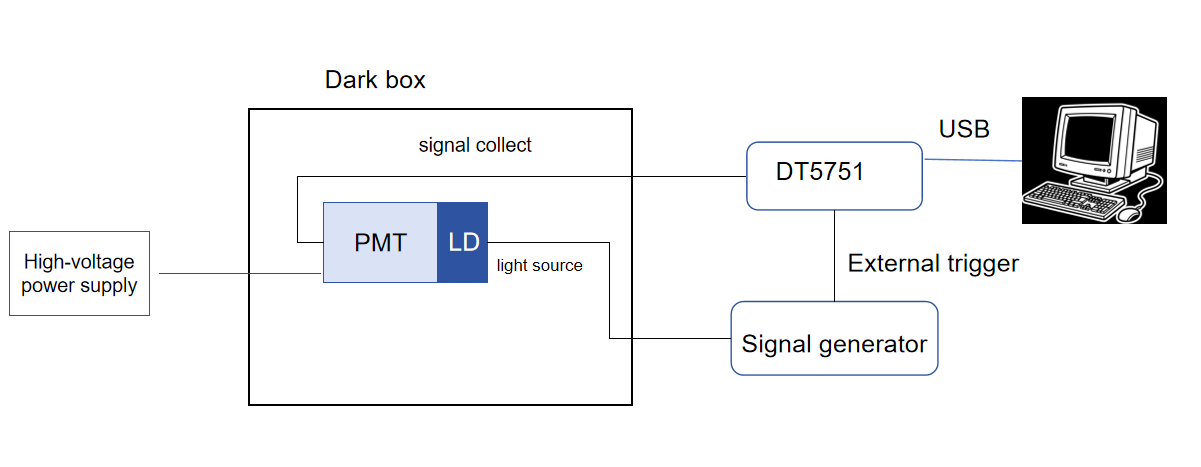}
	\caption{Schematic of the PMT gain calibration setup.
		\label{fig:SPEZZT}}
\end{figure}

As shown in Figure~\ref{fig:SPEZZT}, an experimental system was constructed for PMT gain calibration. The setup consisted of a PMT, a high-voltage power supply, a signal generator, a single-photon light source based on a laser diode (LD), and a CAEN DT5751 digitizer module with a bandwidth of 500~MHz and a sampling rate of 1~GS/s~\cite{CAEN:DT5751}. 
By adjusting the high-voltage power supply, the bias voltage applied to the PMT was precisely controlled, enabling accurate gain calibration.

\begin{figure}[H]
	\includegraphics[width=12.0 cm]{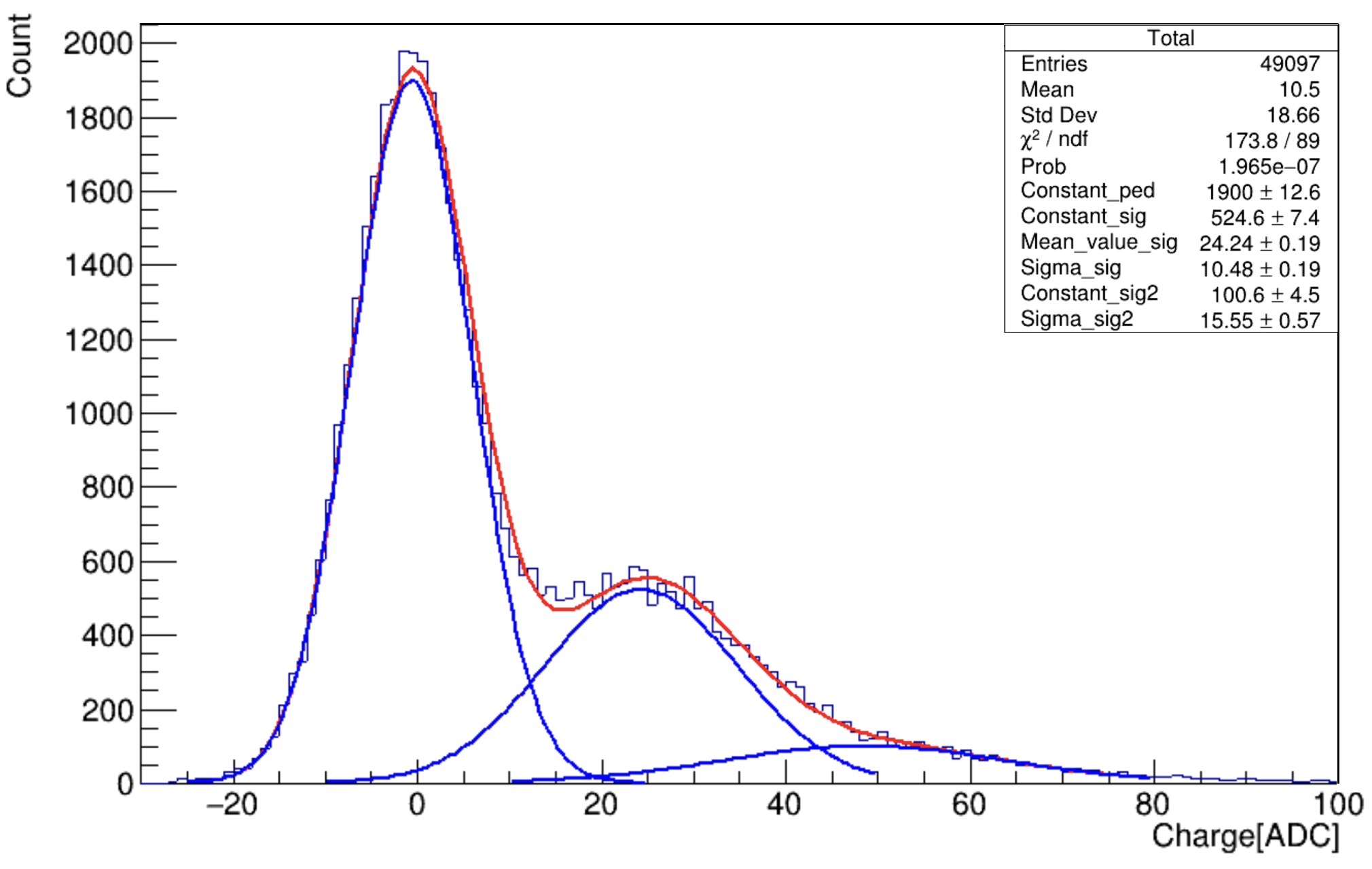}
	\caption{Charge-integrated spectrum of SPE signals.
		\label{fig:PMTgain}}
\end{figure}

The PMT bias voltage was scanned from $-1200$ to $-1400$~V in steps of 50~V. For single-photoelectron (SPE) measurements, the settings of the signal generator for driving the LD were first adjusted iteratively while monitoring the PMT output on the CAEN DT5751 digitizer. After repeated tuning, low-amplitude PMT pulses corresponding to the SPE regime were observed. Under these conditions, the signal generator was operated in the pulse mode with continuous output, and the pulse frequency, pulse width, and output amplitude were finally set to 2~kHz, 100~ns, and 1.36~V, respectively. The LD was then used as the light source for SPE calibration. The LD-induced signal pulses were charge-integrated, and the resulting integrals were filled into a histogram. The charge spectrum obtained under these conditions was taken as the SPE spectrum. The integrated for charge SPE spectrum is shown in Figure~\ref{fig:PMTgain} and was fitted with the sum of three Gaussian functions.

\begin{figure}[H]
	\includegraphics[width=12.0 cm]{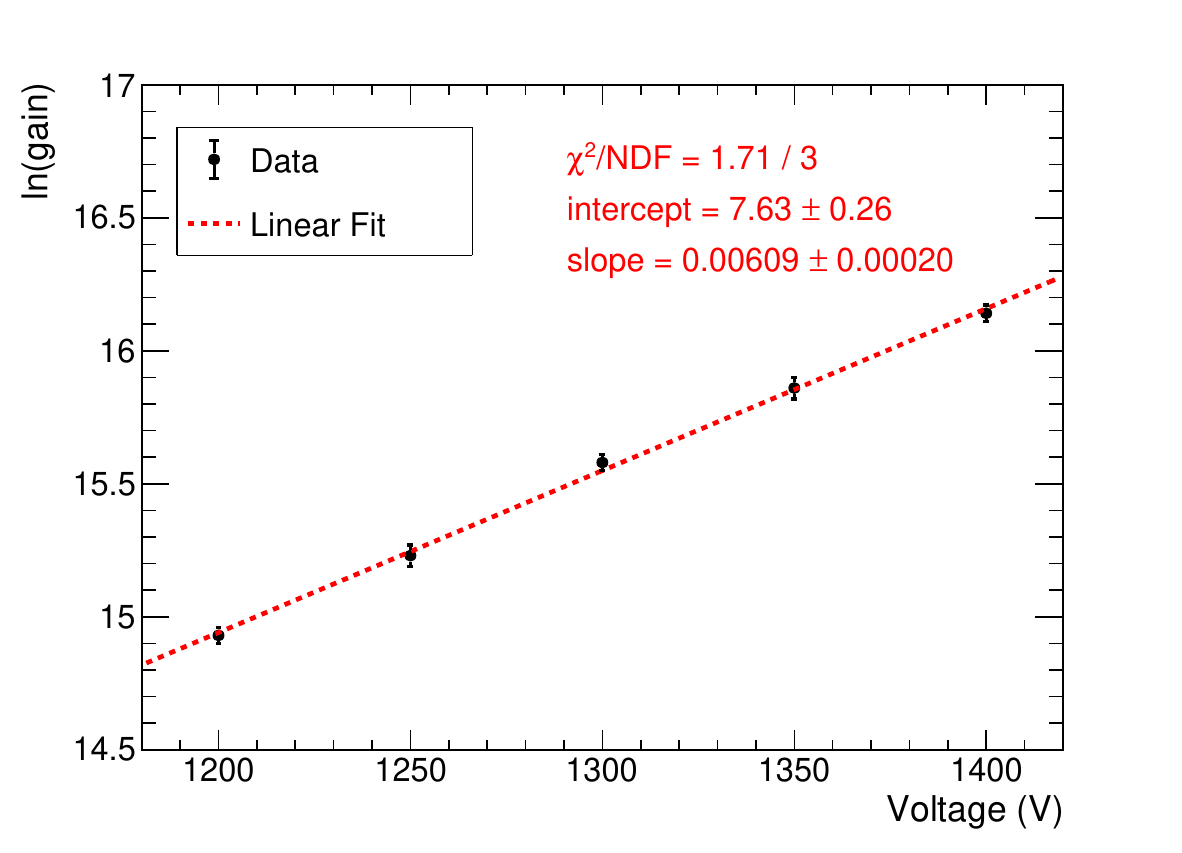}
	\caption{Relationship between the logarithm of PMT gain and applied voltage. The error bars represent the fitting uncertainties of the mean values obtained from Gaussian fits.
		\label{fig:Gain-vol}}
\end{figure}

The first peak corresponds to the pedestal, the second is the SPE peak, and the third is the two-photoelectron (2PE) peak. After identifying the operating voltage corresponding to the SPE response, the PMT bias voltage was varied stepwise to establish the voltage--gain relationship, as also shown in Figure~\ref{fig:Gain-vol}. With the PMT biased at $-1200$~V, the gain was measured to be approximately $3.0 \times 10^{6}$. The corresponding SPE pulse integral was 24.24 in DT5751 channel units, which corresponds to 23.67~mV\,ns after conversion using a factor of $1000/1024$. The integration window used for SPE charge extraction was 100~ns.

\subsection{Detector Configurations and Measurement Scheme}
\label{sec:2.3}
Two detector configurations were employed in this study. 

(1) EJ200 + EJ426 + PMT configuration: This configuration was used for $\gamma$-response calibration and for evaluating thermal-neutron/$\gamma$ discrimination performance. EJ200 provided the $\gamma$-equivalent energy response, whereas EJ426 served as the thermal-neutron-sensitive layer.

(2) EJ276 + EJ426 + PMT configuration: This configuration was primarily used under Am--Be irradiation to investigate pulse shape discrimination among fast neutrons, thermal neutrons, and $\gamma$ rays. EJ276 provided fast-neutron/$\gamma$ discrimination capability, while EJ426 remained sensitive to thermal neutrons.
Gamma-response calibration and neutron/$\gamma$ discrimination measurements were performed for both configurations using three $\gamma$-ray sources ($^{137}$Cs, $^{22}$Na, and $^{60}$Co) and one Am--Be neutron source. During the $\gamma$-response calibration measurements, the activities of the $^{137}$Cs, $^{22}$Na, and $^{60}$Co sources were 1.08$\times$10$^5$~Bq, 1.02$\times$10$^5$~Bq, and 3.8$\times$10$^5$~Bq, respectively. The $\gamma$-ray sources were positioned at a distance of 10~cm from the source-facing surface of the detector assembly, and the PMT bias voltage and acquisition time were set to  $-1200$~V and 600~s, respectively. The Am--Be neutron source used in this study had a neutron emission rate of approximately $10^{6}$~n/s. During the Am--Be measurements, the source-to-detector distance was 10~cm, and the PMT bias voltage and acquisition time were $-1200$~V and 600~s, respectively. HDPE moderators with thicknesses ranging from 1~cm to 3~cm were used to obtain different neutron moderation conditions.

These two configurations were used to assess the mixed-field response of the detector assemblies under the tested conditions.

\subsection{Gamma Energy Calibration}
\label{sec:2.4}
To enable a consistent comparison of detector responses, a gamma energy calibration was performed for the EJ200+EJ426 and EJ276+EJ426 assemblies. Three standard gamma-ray sources, $^{137}$Cs, $^{22}$Na, and $^{60}$Co, were used to provide calibration points over the relevant energy range. The full-energy peaks and corresponding Compton-edge energies of these calibration sources are listed in Table~\ref{tab:gamma_sources}.
\begin{table}[H]
	\caption{Full-energy peaks and corresponding Compton-edge energies of the three calibration gamma-ray sources.\label{tab:gamma_sources}}
	\begin{tabularx}{\textwidth}{CCC}
		\toprule
		\textbf{Gamma-ray source} & \textbf{$E_{\mathrm{peak}}$ (MeV)} & \textbf{$E_{\mathrm{Compton}}$ (MeV)} \\
		\midrule
		$^{137}$Cs & 0.662 & 0.477 \\
		$^{22}$Na  & 0.511; 1.275 & 0.341; 1.062 \\
		$^{60}$Co  & 1.173; 1.333 & 0.963; 1.118 \\
		\bottomrule
	\end{tabularx}
\end{table}

The scintillators used in this study are organic materials composed primarily of carbon and hydrogen. Owing to their relatively low density and low effective atomic number, their gamma-ray stopping power is limited, and full-energy absorption peaks are not expected to be prominent in the measured spectra. In the energy range relevant to this study, gamma interactions in the scintillator are dominated mainly by Compton scattering. Therefore, the gamma energy calibration was performed using the Compton edges of the measured spectra. 
 
To obtain the gamma-response spectra for the two plastic-scintillator assemblies, measurements were performed with the oscilloscope over an acquisition period of 600~s with a threshold of 40~mV. For each source, signals were recorded both with the source present and without the source, and the net energy-response spectrum was obtained by subtracting the background spectrum from the source spectrum.

For $^{60}$Co, the two gamma-ray energies are close to each other and could not be resolved separately because of the limited energy resolution of this setup. Therefore, the mean value of the two corresponding Compton-edge energies was used as the calibration point for the linear fit. 
 
As described in Section~\ref{sec:2.2}, the signal integral corresponding to a SPE was measured to be 24.24 in DT5751 channel units, corresponding to 23.67~mV\,ns, at a PMT bias voltage of $-1200$~V. The Compton-edge positions of the $^{137}$Cs, $^{22}$Na, and $^{60}$Co spectra were determined from the pulse-integrated spectra using a Gaussian-based fit applied to the high-energy edge region. The fitted peak position $\mu$ was then used as the calibration point, and the corresponding fitting uncertainty was taken from the fit. These pulse-integral values were subsequently converted into PE counts using the SPE integral. The relationship between PE yield and gamma energy was then fitted with a linear equation 
 \begin{equation} N_{\mathrm{PE}} = p_{1} E + p_{0}, \label{eq:gamma_calibration} \end{equation} 
where $p_{1}$ represents the light yield in PE/MeV and $p_{0}$ accounts for residual offsets introduced by the readout and signal-processing chain. Within the calibrated energy range, both detector assemblies exhibited an approximately linear response. 
 
This calibration relation was subsequently used to convert detector signals into gamma energy for the energy-resolved pulse shape discrimination analysis presented in the following section.
\subsection{Pulse Shape Discrimination Method}
\label{sec:2.5}

\begin{figure}[H]
	\centering
	\subfloat[EJ276 signal waveform]{
		\includegraphics[width=0.48\textwidth]{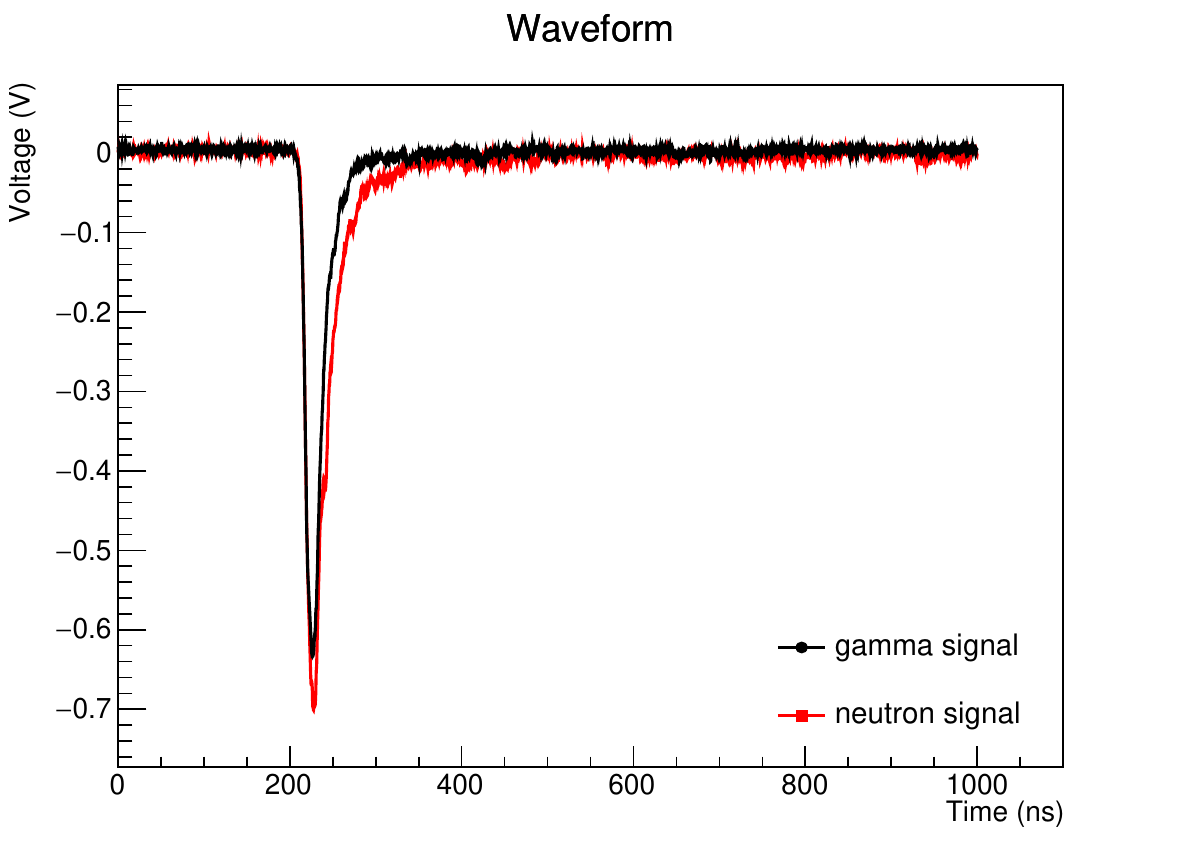}
	}
	\hfill
	\subfloat[EJ426 signal waveform]{
		\includegraphics[width=0.48\textwidth]{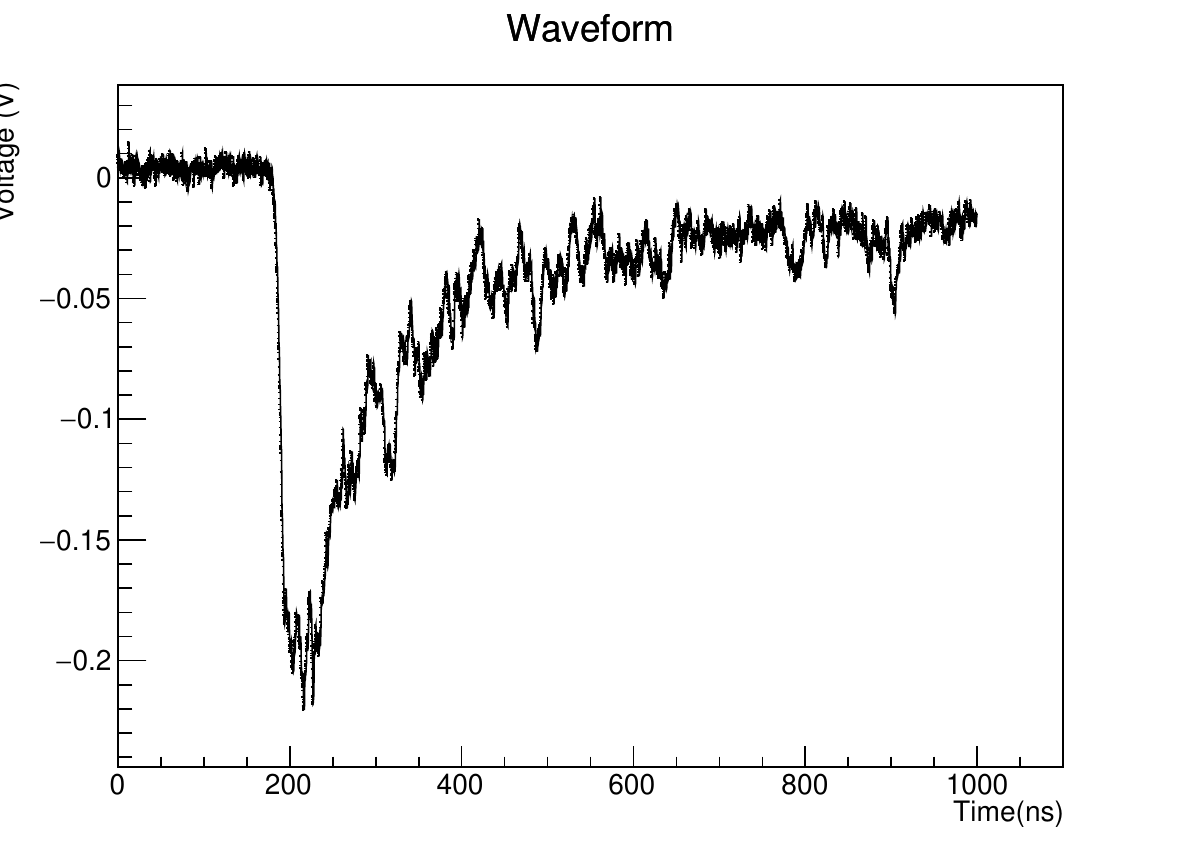}
	}
   \caption{Representative signal waveforms measured from the EJ276 and EJ426 scintillators.}

	\label{fig:Signal_Waveform}
\end{figure}

Pulse shape discrimination (PSD) measurements were performed using the EJ200+EJ426 and EJ276+EJ426 detector assemblies under irradiation with the Am--Be neutron source. The recorded waveforms were analyzed offline to distinguish $\gamma$-ray, fast neutron, and thermal-neutron events on the basis of their different scintillation decay characteristics. The representative signal waveforms of EJ276 and EJ426 are shown in Figure~\ref{fig:Signal_Waveform}, demonstrating the faster pulse response of EJ276 and the slower scintillation component of EJ426. In the composite detector assemblies, the EJ426 screen produced relatively slow pulses associated with thermal-neutron capture, whereas the plastic    scintillators exhibited faster scintillation responses. In particular, EJ276 enabled pulse-shape-based discrimination between fast neutrons and   $\gamma$ rays, while EJ200 mainly contributed to the prompt scintillation signal without intrinsic fast neutron/  $\gamma$ PSD capability.
For each event, the pulse integral was calculated over two different time windows after the trigger. A short integration gate of 75~ns was used to characterize the prompt component of the signal, whereas a long integration gate of 800~ns was used to represent the total collected charge. The pulse-shape parameter was then defined as
\begin{equation}
	PSD = 1 - \frac{Q_{\mathrm{short}}}{Q_{\mathrm{long}}},
	\label{eq:psd_parameter}
\end{equation}

where $Q_{\mathrm{short}}$ and $Q_{\mathrm{long}}$ denote the charge integrals obtained within the short and long gates, respectively. With this definition, signals containing a larger slow component exhibit higher PSD values.

Two-dimensional distributions of PSD versus pulse integral were constructed for the different detector assemblies and moderator conditions. One-dimensional projections were then obtained for PSD peak identification and fitting. For the EJ200+EJ426 assembly, the analysis focused mainly on the separation between thermal-neutron and $\gamma$-ray events. For the EJ276+EJ426 assembly, the analysis was extended to the discrimination among fast neutrons, thermal neutrons, and $\gamma$ rays.

The discrimination performance was quantified using the figure of merit (FOM), defined as~\cite{Winyard1971,ref-ej276-calibration}
\begin{equation}
	FOM = \frac{|\mu_{1}-\mu_{2}|}{\mathrm{FWHM}_{1}+\mathrm{FWHM}_{2}},
	\label{eq:fom_definition}
\end{equation}
where $\mu_{1}$ and $\mu_{2}$ are the central positions of the two PSD peaks under consideration, and $\mathrm{FWHM}_{1}$ and $\mathrm{FWHM}_{2}$ are their full widths at half maximum. For the EJ276+EJ426 assembly, separate FOM values were evaluated for the fast-neutron/$\gamma$, thermal-neutron/$\gamma$, and thermal-neutron/fast neutron peak pairs whenever the corresponding peaks could be resolved.

To investigate the energy dependence of the discrimination performance, the pulse-integral values were converted into gamma energy using the calibration relation established in Section~\ref{sec:2.4}. It should be noted that the neutron energy was not calibrated, therefore, the energies of the  neutron events in the PSD plot are actually equivalent gamma energies. The data were then analyzed within selected gamma-energy intervals using 500~keV windows, and the corresponding PSD projections were fitted to evaluate the variation of the FOM with energy. This procedure was used in particular to assess the energy range over which fast-neutron/$\gamma$ discrimination in the EJ276+EJ426 assembly became effective.

For the Am--Be measurements, HDPE moderators with different thicknesses were introduced on the source-facing side of the detector assembly to study the effect of neutron moderation on the PSD distributions. The same PSD procedure was applied to all moderator conditions so that the relative changes in thermal-neutron, fast neutron, and $\gamma$-ray event populations could be compared consistently.

\section{Results}
\subsection{Energy Calibration Results}

\begin{figure}[H]
	\centering
	\subfloat[$^{137}$Cs source]{
		\includegraphics[width=0.48\textwidth]{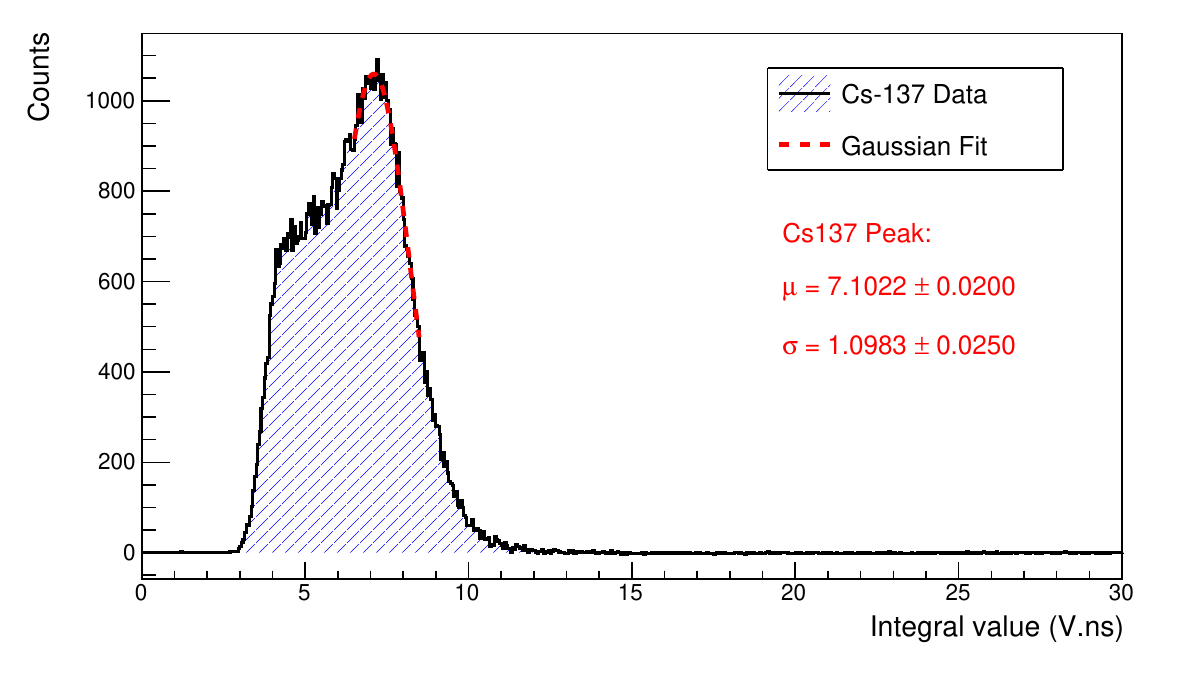}
	}
	\hfill
	\subfloat[$^{22}$Na source]{
		\includegraphics[width=0.48\textwidth]{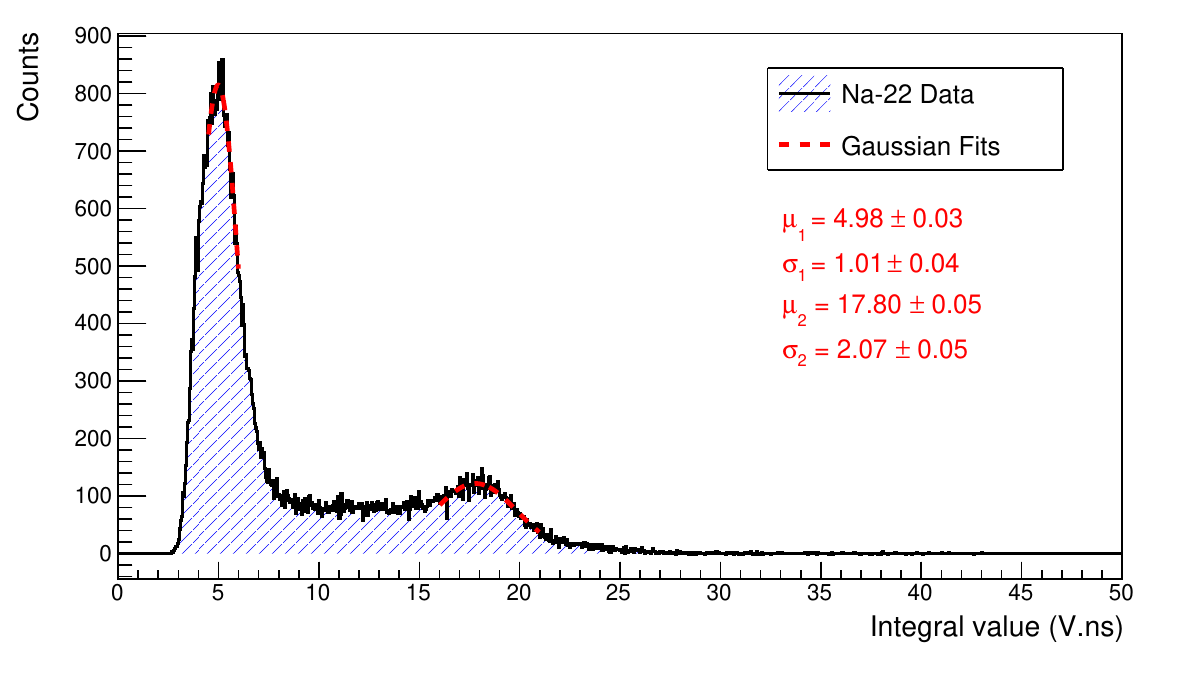}
	}
	
	\vspace{0.5em}
	
	\subfloat[$^{60}$Co source]{
		\includegraphics[width=0.48\textwidth]{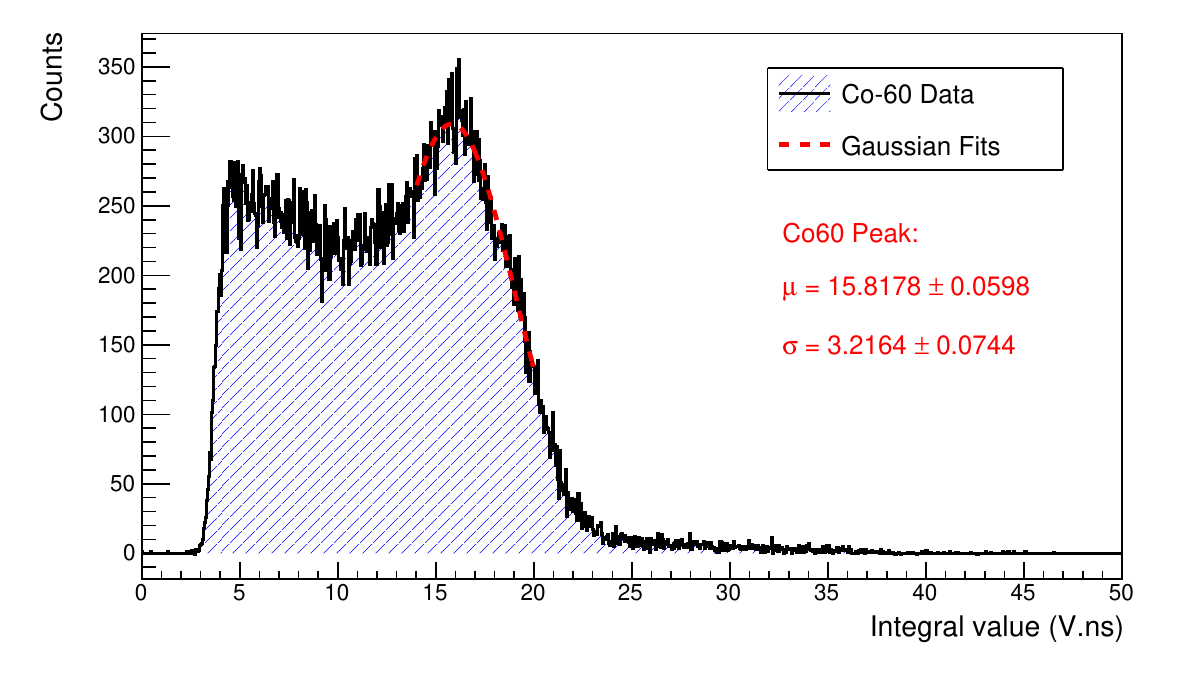}
	}
	\caption{Pulse-integrated gamma-response spectra of the EJ200+EJ426 assembly measured with the $^{137}$Cs, $^{22}$Na, and $^{60}$Co sources. The Compton-edge regions were fitted to extract the corresponding calibration points.}
	\label{fig:Comptonfit}
\end{figure}

The gamma-response spectra obtained from the EJ200+EJ426 and EJ276+EJ426 assemblies were analyzed using the calibration procedure described in Section~\ref{sec:2.4}. After background subtraction, the Compton-edge regions of the $^{137}$Cs, $^{22}$Na, and $^{60}$Co spectra were identified and fitted to extract the corresponding calibration points. Representative fitted spectra are shown in Figure~\ref{fig:Comptonfit}.

\begin{figure}[H]
	\centering
	\subfloat[EJ200+EJ426 assembly]{
		\includegraphics[width=0.48\textwidth]{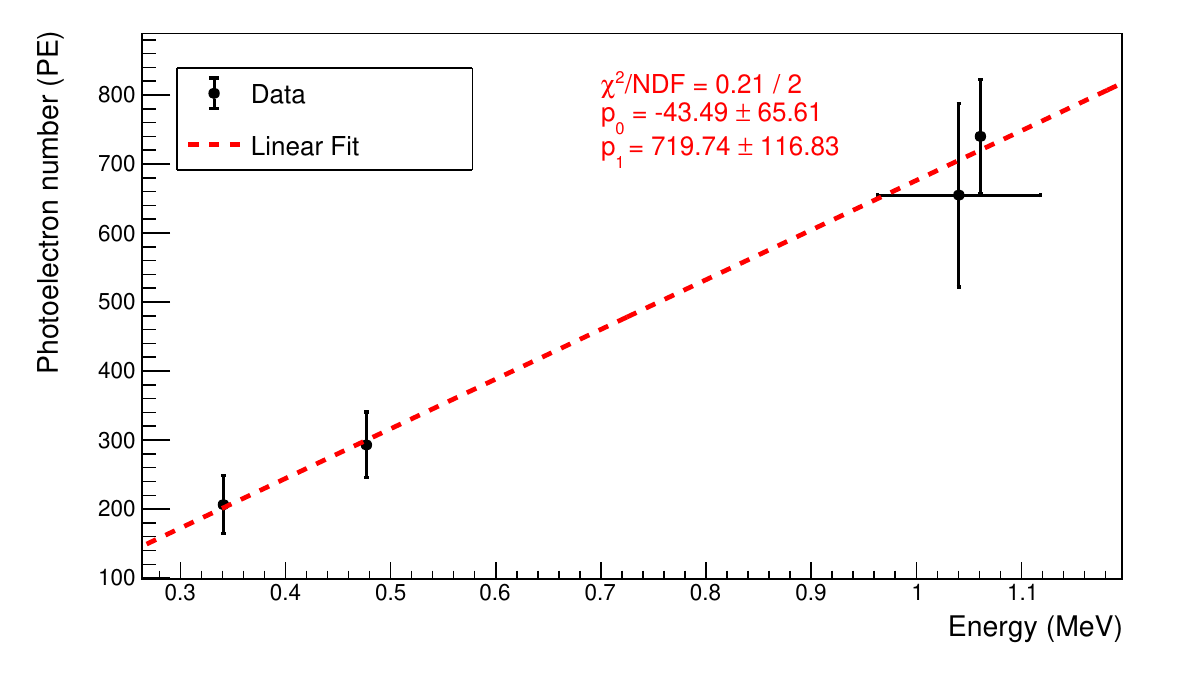}
	}
	\hfill
	\subfloat[EJ276+EJ426 assembly]{
		\includegraphics[width=0.48\textwidth]{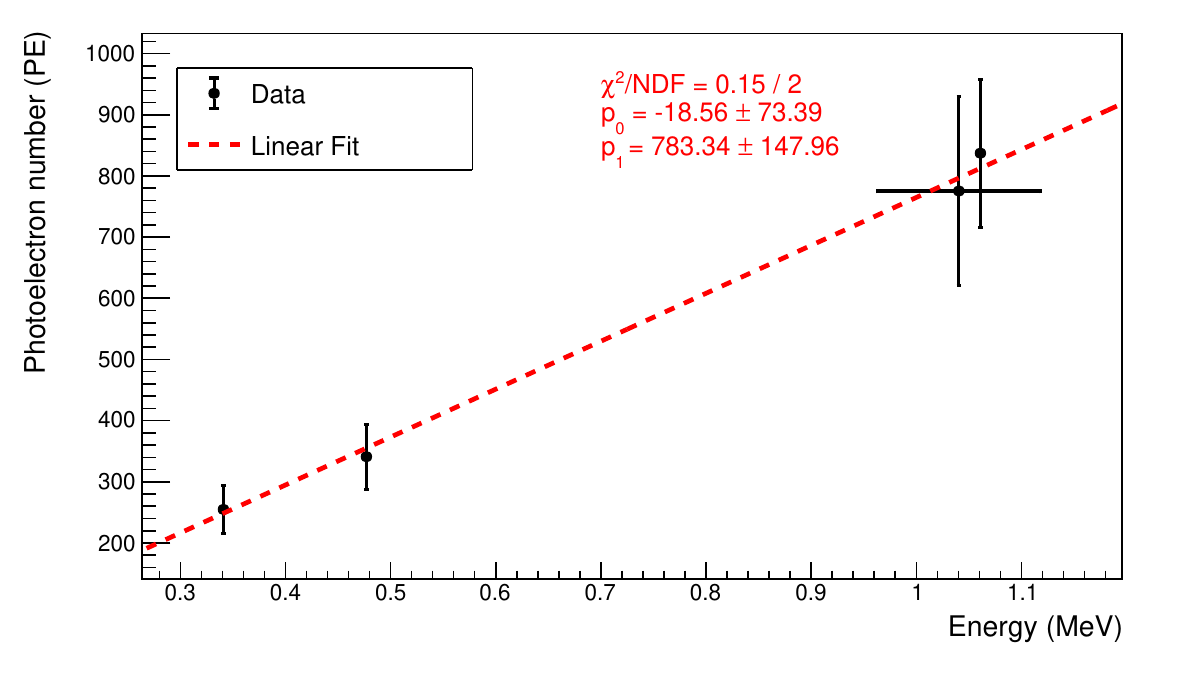}
	}
	\caption{Linear calibration results for the two detector assemblies. (a) Relationship between PE yield and gamma-equivalent energy for the EJ200+EJ426 assembly. (b) Relationship between PE yield and gamma-equivalent energy for the EJ276+EJ426 assembly. The solid lines represent the corresponding linear fits.}
	\label{fig:linearity}
\end{figure}

For both detector assemblies, the pulse integrals were converted into PE yields using the SPE integral determined in Section~\ref{sec:2.2}, and the resulting PE yields were fitted with a linear function as a function of gamma energy. The fitted linear relationships are shown in Figure~\ref{fig:linearity}.

Within the calibrated energy range, both detector assemblies exhibited a linear response. This linearity provides the basis for the subsequent energy-resolved PSD analysis. For the EJ200+EJ426 assembly, the fitted parameters were $p_{1}=716.5 \pm 114.2$~PE/MeV and $p_{0}=-42.19 \pm 64.82$, whereas for the EJ276+EJ426 assembly, the fitted parameters were $p_{1}=781.9 \pm 144.5$~PE/MeV and $p_{0}=-17.99 \pm 72.34$. The corresponding fitting uncertainties were obtained from the linear fit.

\subsection{Pulse shape discrimination of the EJ200+EJ426 Assembly}

\begin{figure}[H]
	\centering
	\subfloat[1~cm HDPE moderator]{
		\includegraphics[width=0.48\textwidth]{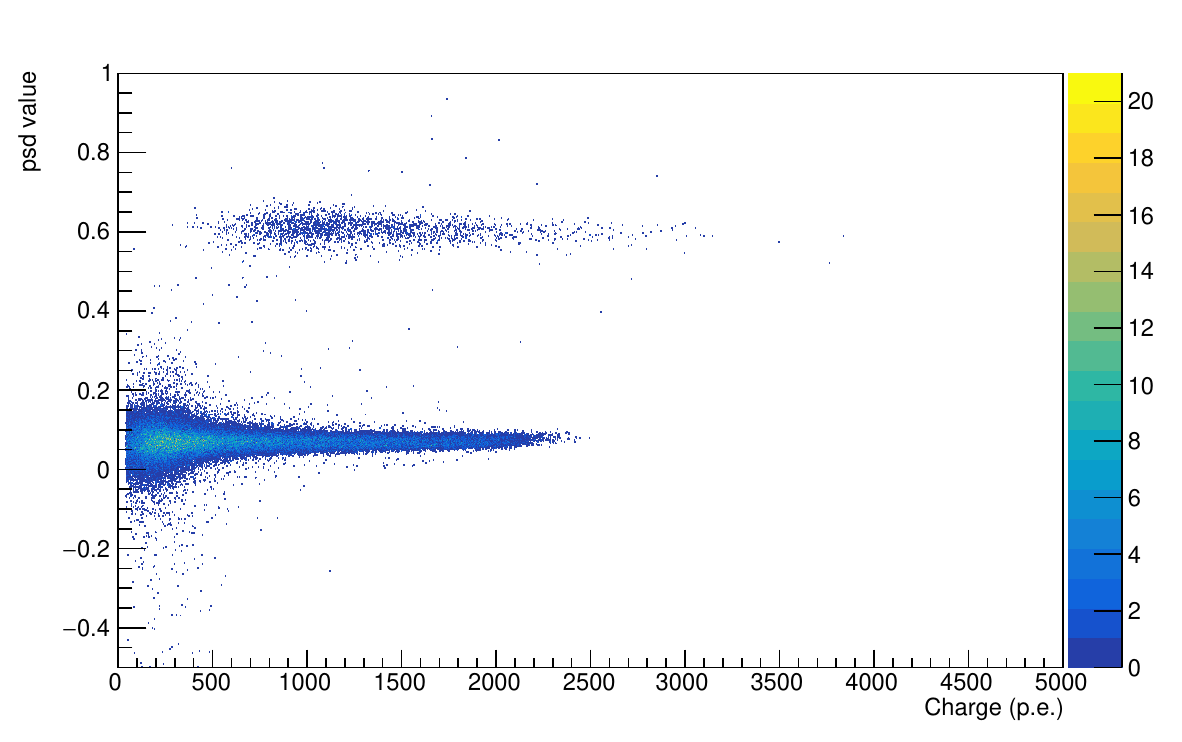}
	}
	\hfill
	\subfloat[3~cm HDPE moderator]{
		\includegraphics[width=0.48\textwidth]{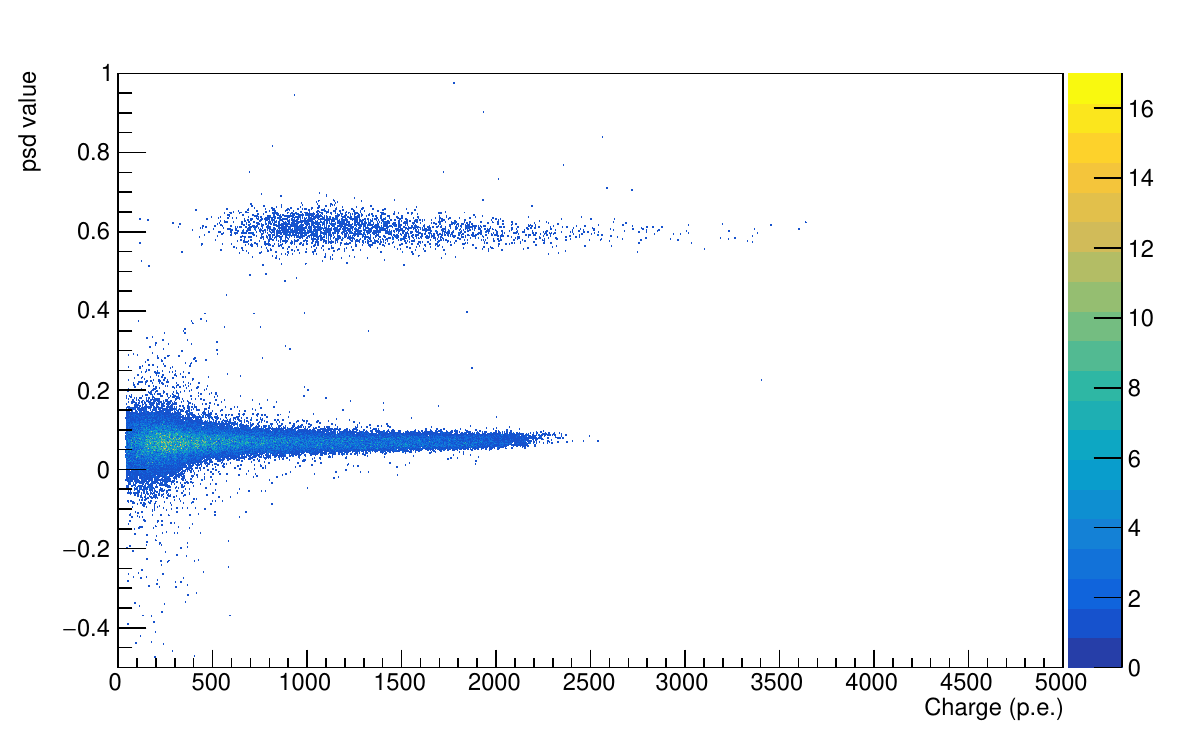}
	}
	\caption{Two-dimensional distributions of PSD value versus charge (PE) for the EJ200+EJ426 assembly under Am--Be irradiation. In both cases, two main event populations can be identified, corresponding predominantly to $\gamma$-ray-induced events and thermal-neutron capture events. The relative population of the high-PSD band increases with increasing HDPE thickness.}
	\label{fig:EJ200PSD2D}
\end{figure}

\begin{figure}[H]
	\centering
	\subfloat[1~cm HDPE moderator]{
		\includegraphics[width=0.48\textwidth]{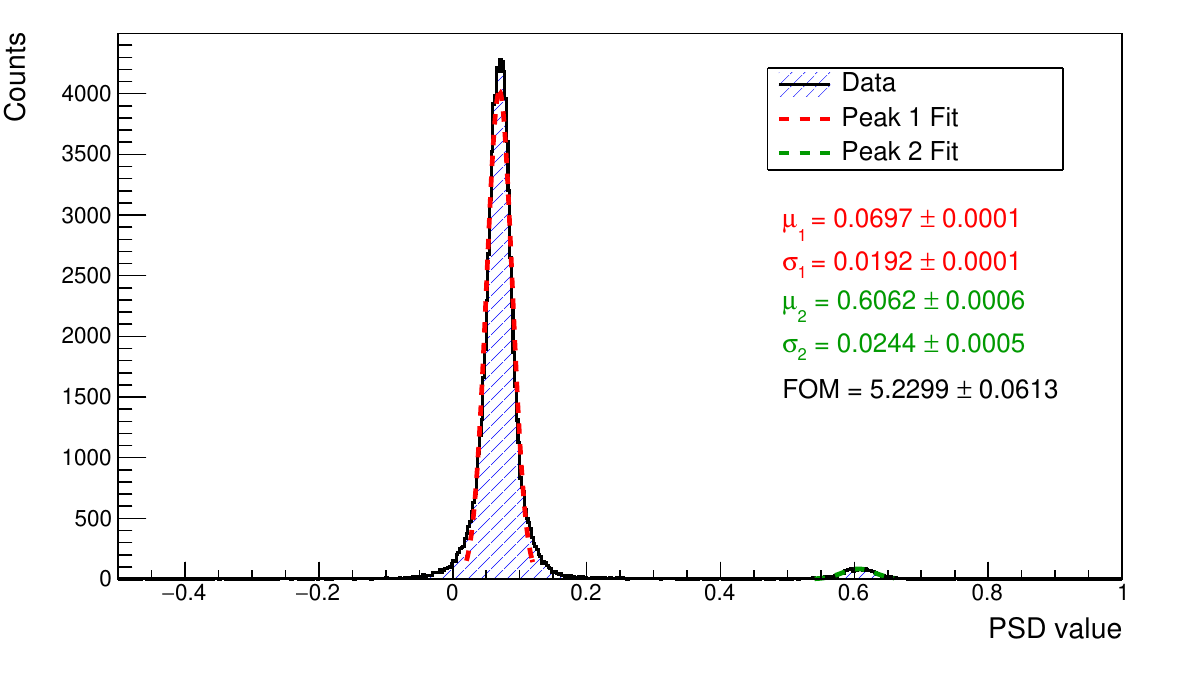}
	}
	\hfill
	\subfloat[3~cm HDPE moderator]{
		\includegraphics[width=0.48\textwidth]{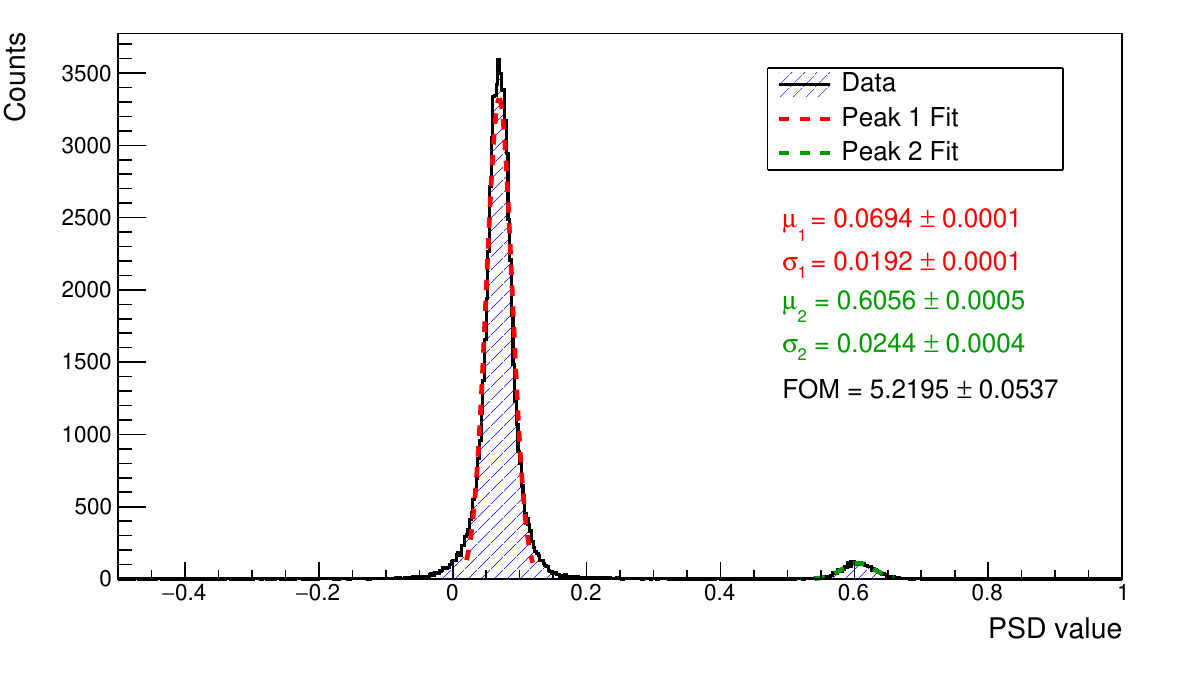}
	}
	\caption{One-dimensional PSD projections and Gaussian fits for the EJ200+EJ426 assembly under Am--Be irradiation. For both HDPE thicknesses shown in the left and right panels, two PSD components were identified and fitted, corresponding mainly to $\gamma$-ray-induced events and thermal-neutron capture events. The extracted FOM values indicate effective thermal-neutron/$\gamma$ discrimination under the tested conditions.}
	\label{fig:EJ200PSD1D}
\end{figure}

The pulse shape discrimination results of the EJ200+EJ426 assembly with different HDPE moderator thicknesses under Am--Be irradiation are shown in Figures~\ref{fig:EJ200PSD2D} and \ref{fig:EJ200PSD1D}. In the two-dimensional distributions of PSD value versus charge, two main event populations can be identified. The low-PSD band is attributed mainly to $\gamma$-ray-induced events in the EJ200 scintillator, whereas the high-PSD band corresponds to thermal-neutron capture events associated with the EJ426 screen. As the HDPE moderator thickness increases from 1~cm to 3~cm, the relative population of the high-PSD band increases, indicating an enhanced moderation of the neutrons.

To quantify the separation performance, one-dimensional PSD projections were extracted and fitted with two Gaussian functions, as shown in Figure~\ref{fig:EJ200PSD1D}. For the 1~cm HDPE condition, the fitted peak positions were $\mu_{1}=0.0697$ and $\mu_{2}=0.6062$, with corresponding widths of $\sigma_{1}=0.0192$ and $\sigma_{2}=0.0244$, yielding an FOM of $5.2299 \pm 0.0613$. For the 3~cm HDPE condition, the fitted peak positions were $\mu_{1}=0.0694$ and $\mu_{2}=0.6056$, with widths of $\sigma_{1}=0.0192$ and $\sigma_{2}=0.0244$, corresponding to an FOM of $5.2195 \pm 0.0537$.

The large separation between the two PSD components and the consistently high FOM values confirm that the EJ200+EJ426 assembly provides effective thermal-neutron/$\gamma$ discrimination under the tested conditions.

\subsection{Pulse shape discrimination of the EJ276+EJ426 Assembly}

\begin{figure}[H]
	\centering
	\subfloat[1~cm HDPE moderator]{
		\includegraphics[width=0.48\textwidth]{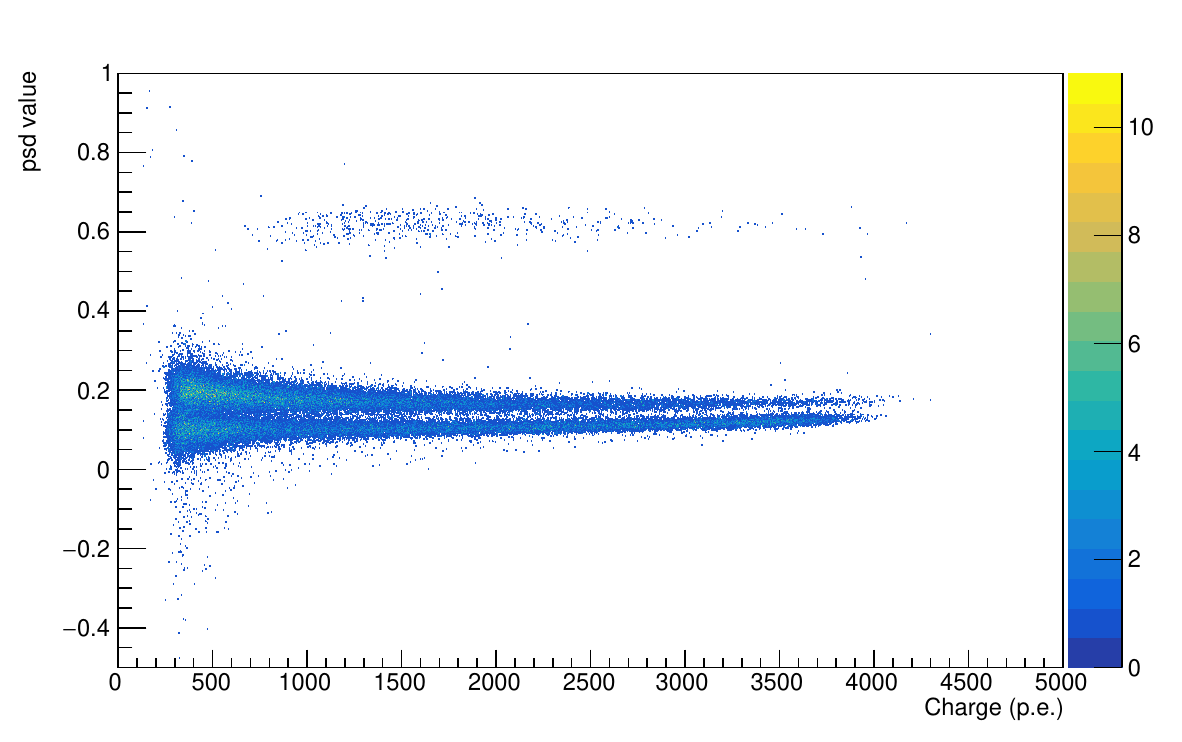}
	}
	\hfill
	\subfloat[3~cm HDPE moderator]{
		\includegraphics[width=0.48\textwidth]{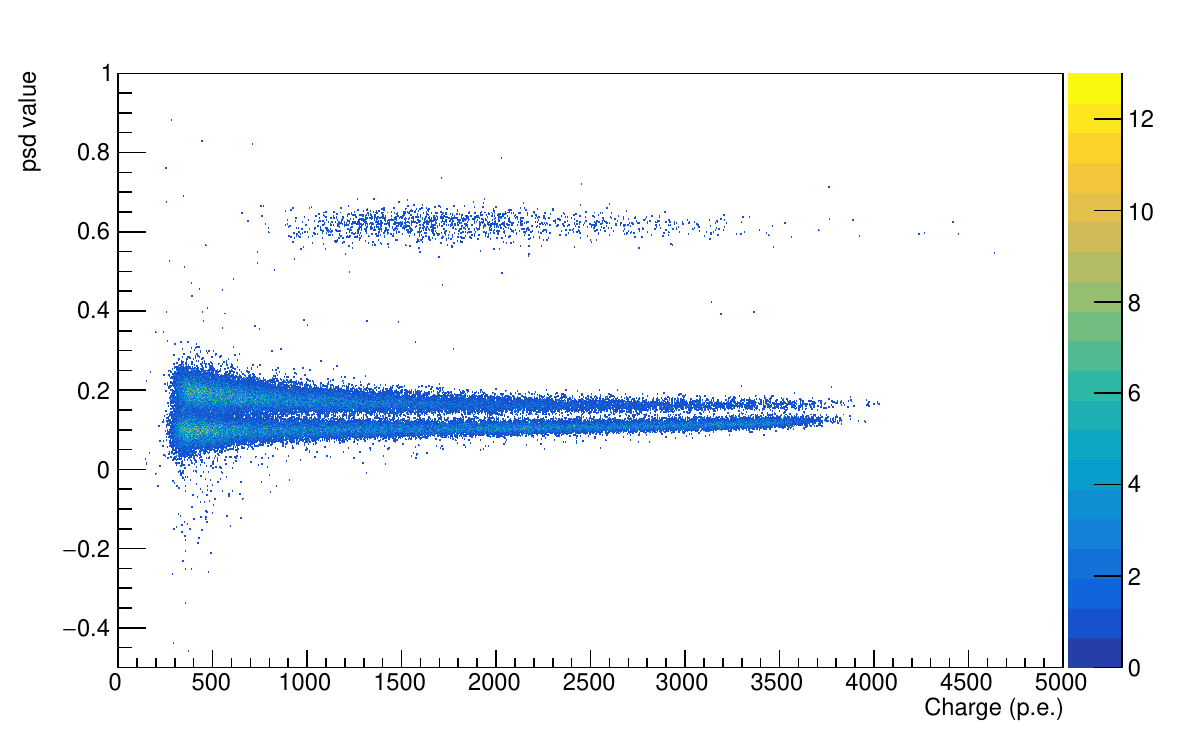}
	}
	\caption{Two-dimensional distributions of PSD value versus charge (PE) for the EJ276+EJ426 assembly under Am--Be irradiation. Left: measurement with a 1~cm HDPE moderator. Right: measurement with a 3~cm HDPE moderator. Three main event populations can be identified, corresponding predominantly to $\gamma$-ray-induced events, fast neutron-induced events, and thermal-neutron capture events.}
	\label{fig:EJ276PSD2D}
\end{figure}

\begin{figure}[H]
	\centering
	\subfloat[1~cm HDPE moderator]{
		\includegraphics[width=0.48\textwidth]{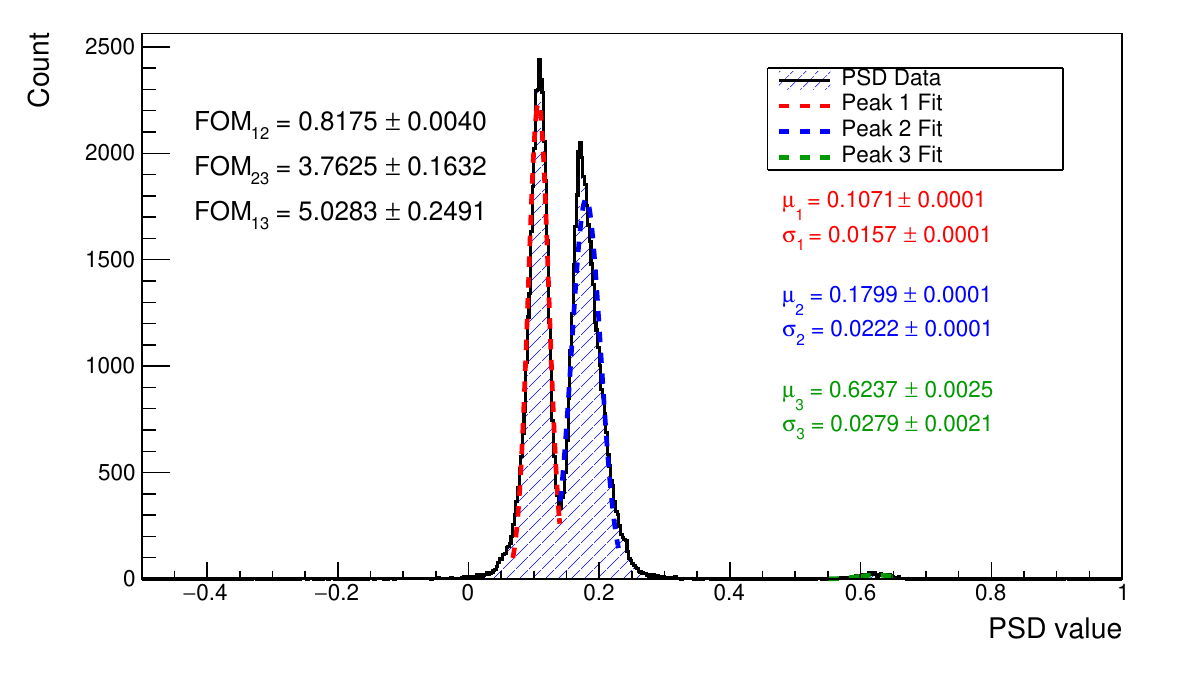}
	}
	\hfill
	\subfloat[3~cm HDPE moderator]{
		\includegraphics[width=0.48\textwidth]{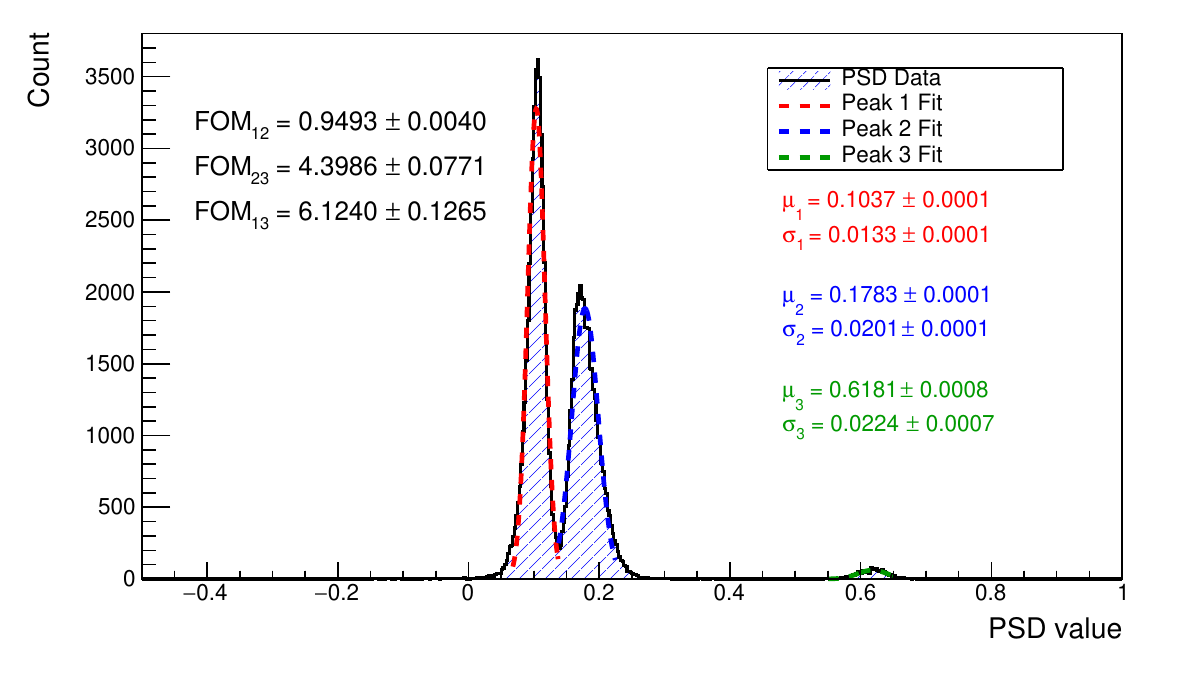}
	}
	\caption{One-dimensional PSD projections and Gaussian fits for the EJ276+EJ426 assembly under Am--Be irradiation. Left: result obtained with a 1~cm HDPE moderator. Right: result obtained with a 3~cm HDPE moderator. Three PSD components were identified and fitted, corresponding mainly to $\gamma$-ray-induced events, fast neutron-induced events, and thermal-neutron capture events.}
	\label{fig:EJ276PSD1D}
\end{figure}

\begin{figure}[H]
	\centering
	\includegraphics[width=0.62\textwidth]{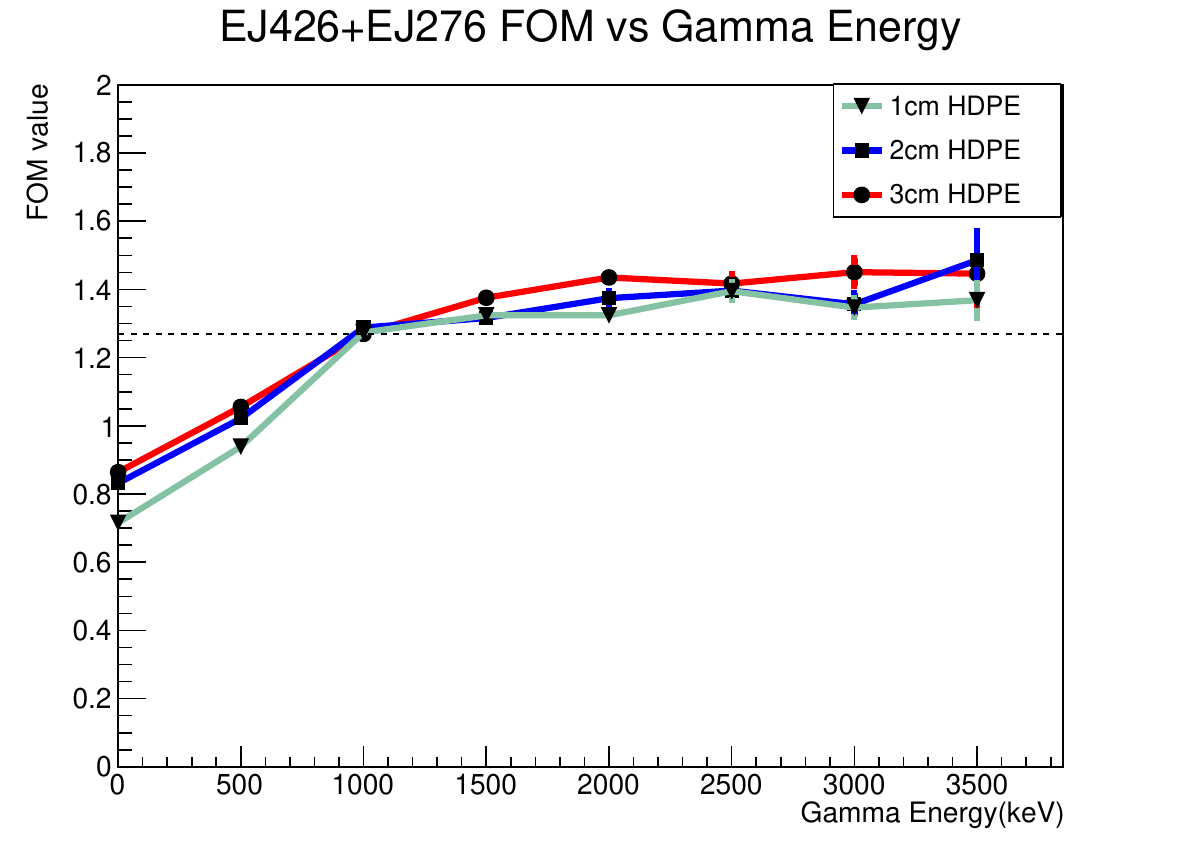}
	\caption{Energy dependence of the fast-neutron/$\gamma$ discrimination performance for the EJ276+EJ426 assembly. The FOM was evaluated in selected gamma-equivalent energy intervals under different HDPE moderator thicknesses. The dashed line indicates FOM = 1.27.}
	\label{fig:EJ276FOMEnergy}
\end{figure}

The pulse shape discrimination results of the EJ276+EJ426 assembly under Am--Be irradiation are shown in Figures~\ref{fig:EJ276PSD2D}, \ref{fig:EJ276PSD1D}, and \ref{fig:EJ276FOMEnergy}. In the two-dimensional distributions of PSD value versus charge, three main event populations can be identified. The low-PSD band is attributed mainly to $\gamma$ events in the EJ276 scintillator, the intermediate-PSD band corresponds predominantly to fast-neutron events which actually are the recoiled protons depositing their energy in the scintillator, and the high-PSD band is associated with thermal-neutron capture events in the EJ426 screen. Compared with the EJ200+EJ426 assembly, the EJ276+EJ426 configuration exhibits an additional PSD component arising from the intrinsic fast-neutron/$\gamma$ discrimination capability of EJ276.

To quantify the separation performance, one-dimensional PSD projections were extracted and fitted with three Gaussian functions, as shown in Figure~\ref{fig:EJ276PSD1D}. For the 1~cm HDPE condition, the fitted peak positions were $\mu_{1}=0.1071$, $\mu_{2}=0.1799$, and $\mu_{3}=0.6237$, with corresponding widths of $\sigma_{1}=0.0157$, $\sigma_{2}=0.0222$, and $\sigma_{3}=0.0279$. The corresponding FOM values were $0.8175 \pm 0.0040$ for the fast-neutron/$\gamma$ pair (FOM$_{12}$), $3.7625 \pm 0.1632$ for the thermal-neutron/fast neutron pair (FOM$_{23}$), and $5.0283 \pm 0.2491$ for the thermal-neutron/$\gamma$ pair (FOM$_{13}$). For the 3~cm HDPE condition, the fitted peak positions were $\mu_{1}=0.1039$, $\mu_{2}=0.1778$, and $\mu_{3}=0.6181$, with widths of $\sigma_{1}=0.0130$, $\sigma_{2}=0.0194$, and $\sigma_{3}=0.0224$. The corresponding FOM values were $0.9678 \pm 0.0043$ for FOM$_{12}$, $4.4766 \pm 0.0798$ for FOM$_{23}$, and $6.1701 \pm 0.1284$ for FOM$_{13}$.

The fitted results show that the separation between the thermal-neutron peak and the other two PSD components is pronounced, whereas the fast neutron and $\gamma$-ray peaks remain partially overlapped in the low-energy range. In both moderator conditions, the thermal-neutron/$\gamma$ and thermal-neutron/fast neutron separations are strong, with FOM values exceeding 3.7, whereas the fast-neutron/$\gamma$ FOM value is below 1.0. These results indicate that, in the full energy spectrum analysis, the EJ276+EJ426 assembly provides clear thermal-neutron tagging and additional fast-neutron/$\gamma$ discrimination capability, although the fast-neutron/$\gamma$ separation is limited in the low-energy range.

To investigate the energy dependence of the fast-neutron/$\gamma$ discrimination performance, the pulse-integral values were converted into gamma-equivalent energy using the calibration relation established in Section~\ref{sec:2.4}. The resulting FOM variation with gamma energy is shown in Figure~\ref{fig:EJ276FOMEnergy}. As the gamma energy increases, the separation between the fast neutron and $\gamma$-ray PSD components becomes more pronounced, and the corresponding FOM increases accordingly. In particular, the fast-neutron/$\gamma$ FOM exceeds the commonly used effective-separation criterion of 1.27 (equivalent to a 3$\sigma$ separation~\cite{Winyard1971,ref-ej276-calibration}) above approximately 1~MeV gamma-equivalent energy.

Under the present experimental conditions, the EJ276+EJ426 assembly provides a practical route toward discrimination among fast neutrons, thermal neutrons, and $\gamma$ rays when the applicable gamma energy range is taken into account. To further examine the effect of neutron moderation, the event populations identified from the PSD analysis were counted for different HDPE moderator thicknesses, as summarized in Table~\ref{tab:hdpe_effect}. With increasing HDPE thickness, the fraction of thermal-neutron events increased from 0.463\% to 0.568\%, whereas the fraction of fast-neutron events decreased from 45.29\% to 38.99\%. Meanwhile, the relative fraction of $\gamma$-ray events increased from 54.24\% to 60.44\%.

\begin{table}[H]
	\caption{Event counts and relative fractions of the three PSD components for the EJ276+EJ426 assembly under different HDPE moderator thicknesses. The data were acquired over 10~min.\label{tab:hdpe_effect}}
	\begin{tabularx}{\textwidth}{CCCCC}
		\toprule
		\textbf{HDPE thickness} & \textbf{Thermal neutrons} & \textbf{Fast neutrons} & \textbf{$\gamma$ rays} & \textbf{Total event count} \\
		\midrule
		1~cm & 674 (0.463\%) & 65\,911 (45.29\%) & 78\,944 (54.24\%) & 145\,529 \\
		2~cm & 709 (0.540\%) & 55\,122 (42.03\%) & 75\,311 (57.43\%) & 131\,142 \\
		3~cm & 672 (0.568\%) & 46\,135 (38.99\%) & 71\,521 (60.44\%) & 118\,328 \\
		\bottomrule
	\end{tabularx}
\end{table}

These results indicate that increasing the HDPE thickness changes the relative event composition observed with the EJ276+EJ426 assembly. A thicker HDPE moderator enhances neutron moderation, leading to a higher fraction of events classified as thermal neutrons and a lower fraction classified as fast neutrons. At larger thicknesses, however, some moderated neutrons may be attenuated within the HDPE through scattering and, to a lesser extent, capture by hydrogen before reaching the detector.

\section{Discussion}

The present results show that the two detector assemblies provide different discrimination characteristics in mixed radiation fields. For the EJ200+EJ426 assembly, the PSD distributions are dominated by two signal populations. The low-PSD component is associated mainly with $\gamma$-ray-induced events in EJ200, whereas the high-PSD component corresponds to thermal-neutron capture events in EJ426. The extracted FOM values exceed 5 under both 1~cm and 3~cm HDPE moderation conditions, indicating stable and effective thermal-neutron/$\gamma$ separation under the tested conditions. Therefore, the EJ200+EJ426 configuration is more appropriately regarded as a compact detector assembly for thermal-neutron tagging relative to the $\gamma$-ray background.

In contrast, the EJ276+EJ426 assembly exhibits three identifiable PSD components that can be associated mainly with $\gamma$ rays, fast neutrons, and thermal neutrons. This behavior is consistent with the intrinsic fast-neutron/$\gamma$ pulse shape discrimination capability of EJ276 and the slow scintillation response of EJ426 for thermal-neutron capture. The fitted results show that the thermal-neutron peak can be well separated from the other two components, whereas the fast neutron and $\gamma$-ray peaks can be clearly separated above 1~MeV gamma energy with an FOM value greater than or equal to 1.27.

The energy-dependent analysis further clarifies the discrimination behavior of the EJ276+EJ426 assembly. As the gamma-equivalent energy increases, the separation between the fast neutron and $\gamma$-ray PSD components becomes more significant, and the corresponding FOM increases accordingly. In particular, the fast-neutron/$\gamma$ FOM exceeds the commonly adopted effective-separation criterion of 1.27  above approximately 1~MeV gamma-equivalent energy~\cite{Winyard1971,ref-ej276-calibration}. 

The moderator-thickness dependence provides additional information for the above interpretation. The absolute thermal-neutron counts do not increase monotonically with moderator thickness, indicating that the present measurements should be interpreted mainly in terms of changes in relative event composition rather than as a direct quantitative measure of neutron thermalization efficiency.

From a detector-design perspective, the present assemblies can be understood as compact phoswich-like configurations in which scintillation components with different temporal characteristics are optically coupled to a common photosensor and distinguished through pulse-shape analysis~\cite{ref-phoswich-review,ref-phoswich-ej276-gs20}. Within this framework, the comparison between EJ200+EJ426 and EJ276+EJ426 shows that the discrimination performance can be tuned according to the measurement objective. The former favors thermal-neutron/$\gamma$ separation, whereas the latter extends the discrimination capability toward fast neutrons when the applicable energy range is taken into account.

Several limitations should also be noted. Firstly, the present work does not yet provide absolute detection efficiencies for fast neutrons, thermal neutrons, and $\gamma$ rays. Secondly, the fast-neutron/$\gamma$ discrimination performance remains limited in the low-energy region. These aspects should be addressed in future work through additional calibration measurements, detector-geometry optimization, and model-supported analysis.

\section{Conclusions}

In this work, two compact plastic-scintillator detector assemblies, EJ200+EJ426 and EJ276+EJ426, were investigated for mixed-field radiation sensing. A gamma-equivalent energy calibration was established using $^{137}$Cs, $^{22}$Na, and $^{60}$Co sources, and the pulse shape discrimination performance was evaluated under Am--Be irradiation with different HDPE moderator thicknesses.

The results show that the EJ200+EJ426 assembly provides effective thermal-neutron/$\gamma$ discrimination under the tested conditions, with FOM values above 5 for both moderator thicknesses considered. The EJ276+EJ426 assembly exhibits three identifiable PSD components associated mainly with $\gamma$ rays, fast neutrons, and thermal neutrons. In the analysis over the full energy range, the thermal-neutron/fast neutron and thermal-neutron/$\gamma$ separations are strong, whereas the fast-neutron/$\gamma$ separation remains limited. The energy-resolved analysis further shows that effective fast-neutron/$\gamma$ discrimination is achieved mainly above approximately 1~MeV gamma-equivalent energy.

These findings indicate that the two detector assemblies are suitable for different but complementary discrimination tasks in mixed radiation fields. The EJ200+EJ426 configuration is more suitable for thermal-neutron tagging relative to the $\gamma$-ray background, whereas the EJ276+EJ426 configuration provides a practical route toward discrimination among fast neutrons, thermal neutrons, and $\gamma$ rays when the applicable gamma-equivalent energy range is considered.

\authorcontributions{Conceptualization, F.A., G.L. and W.W.; methodology, Y.L., F.A. and G.L.; investigation, Y.L. and F.A.; formal analysis, Y.L. and F.A.; visualization, G.L.; writing---original draft preparation, Y.L. and F.A.; writing---review and editing, G.L. and W.W.; supervision, W.W.; discussion, X.Z., D.L. and X.Y. All authors have read and agreed to the published version of the manuscript.}

\funding{This research was supported by the China Postdoctoral Science Foundation (Grant No. 2025M783462), the Fundamental Research Funds for the Central Universities (Grant No. 24qnpy109), and the National Natural Science Foundation of China (Grant No. 12075087).}

\institutionalreview{Not applicable.}

\informedconsent{Not applicable.}

\dataavailability{The data supporting the findings of this study are available from the corresponding author upon reasonable request.}

\acknowledgments{The authors thank the School of Physics and Institut Franco-Chinois de l'Énergie Nucléaire at Sun Yat-sen University for their support. The authors are also grateful to Prof. Yuehuan Wei, Prof. Bo Mei, and Prof. Tao Xiong for their valuable support.}

\conflictsofinterest{The authors declare no conflicts of interest. The funders had no role in the design of the study; in the collection, analyses, or interpretation of data; in the writing of the manuscript; or in the decision to publish the results.}






\begin{adjustwidth}{-\extralength}{0cm}

\reftitle{References}


\bibliography{refs}



%


\PublishersNote{}
\end{adjustwidth}
\end{document}